\documentclass[article]{elsarticle}

\usepackage{lineno,hyperref}
\modulolinenumbers[5]

\journal{Journal of \LaTeX\ Templates}
 \usepackage{graphicx}
\usepackage{dcolumn}
\usepackage{bm}
\usepackage{hyperref}

\usepackage{graphicx}
\usepackage{dcolumn}
\usepackage{bm}
\usepackage{hyperref}

\usepackage{amsmath}
\usepackage[caption=false]{subfig}
\usepackage{gensymb}

\newcommand{\be}{\begin{eqnarray}}
\newcommand{\ee}{\end{eqnarray}}
\newcommand{\beq}{\begin{equation}}
\newcommand{\eeq}{\end{equation}}
\newcommand{\exclude}[1]{}

\graphicspath{{./Figures/}} 

\begin{document}

\begin{frontmatter}
    
       \title{The mysterious diffuse UV   radiation   and    Axion Quark Nugget dark matter model}
       \author{Ariel  Zhitnitsky}
       \address{Department of Physics and Astronomy, University of British Columbia, Vancouver, V6T 1Z1, BC, Canada}
     
      \begin{abstract}
     It has been recently argued \cite{Henry_2014,Akshaya_2018,2019MNRAS.489.1120A} that there is a strong component of the diffuse far-ultraviolet (FUV) background 
     which is hard  to explain by conventional physics in terms of  the dust-scattered starlight.  We propose that  this excess in FUV radiation    might be result of the 
    dark matter annihilation events  within  the so-called axion quark nugget (AQN) dark matter 
    model, which was originally invented for completely different purpose to explain the observed  similarity between the dark   and the visible components  in the Universe, i.e. $\Omega_{\rm DM}\sim \Omega_{\rm visible}$. We  support this proposal by demonstrating that   intensity and the spectral features of the   AQN induced emissions    are consistent with  the corresponding  characteristics of the observed excess of the FUV radiation. If the future studies confirm the puzzling characteristics  observed in  \cite{Henry_2014,Akshaya_2018,2019MNRAS.489.1120A}
it might unlock a much more deeper and fundamental problem of modern  cosmology-- it may reveal the nature of the dark matter. 
   
     \end{abstract}

\begin{keyword}
dark matter, axion, diffuse UV   radiation
 \end{keyword}

\end{frontmatter}


\section{ Introduction and Motivation} 
The title of this work seemingly includes two contradicting terms: the first one is  ``the   radiation", while the second term of the title is  ``the dark matter" (DM)  which, by definition, must decouple  from the baryons and the radiation. However, the main goal of this work is to argue that the puzzling  features observed in  \cite{Henry_2014,Akshaya_2018,2019MNRAS.489.1120A}  in fact  can be explained within   a specific dark matter  model, the  so-called axion quark nugget (AQN) dark matter model  \cite{Zhitnitsky:2002qa}. Therefore,  the  contradiction in the title is only apparent, while the DM in form of the AQNs strongly couples to the baryons and  can emit light.
Furthermore, if the future and more precise analysis  confirms the puzzling observations  \cite{Henry_2014,Akshaya_2018,2019MNRAS.489.1120A} it could serve as an extraordinary evidence revealing the nature of the dark matter, see few additional comments   at the very end of this Introduction.   

Before we proceed with our  arguments we should, first of all,  highlight the mysterious properties  of the observations \cite{Henry_2014,Akshaya_2018,2019MNRAS.489.1120A} which are very hard to understand if interpreted in terms of the conventional astrophysical  phenomena.  
 Indeed,   the widely accepted picture is that  the dominant source of the diffused ultraviolet (UV) background radiation is the dust-scattered radiation of the UV emitting stars.  However, the recent analysis  carried out in \cite{Henry_2014} and  in subsequent  papers 
   \cite{Akshaya_2018,2019MNRAS.489.1120A} disproves   this conventional picture.  The corresponding  refutal arguments are based on the following puzzling features of the observations: 
   
  {  {\bf 1}. ``uniformity puzzle":}  The diffuse radiation is very uniform in both hemispheres, see Figs 7-10 in \cite{Henry_2014}. This feature should be contrasted to the strong non-uniformity in distribution of the UV emitting stars; 
  
   {  {\bf 2}. ``galactic longitude puzzle":}  The diffuse radiation is almost entirely independent of Galactic longitude. This feature must be contrasted with localization of the brightest UV emitting stars which are overwhelmingly confined to the longitude range $180^0-360^0$, which strongly suggests that the diffuse background radiation can hardly  originate in dust-scattered starlight;
   
      { {\bf 3}. ``galactic latitude puzzle":}    The diffuse radiation increases in brightness toward lower     galactic latitude at all galactic longitudes. 
      This should be contrasted with conventional modelling  \cite{Henry_2014} which predicts very low brightness at low Galactic latitudes. 
  This observation  also  suggests that the  recorded  diffused  emission has galactic, not   extragalactic origin. 
  Indeed, the extragalactic light cannot strongly vary  toward lower     galactic latitude. 
   Due to the same reasons,  the observations  also conclusively suggest that this emission has  celestial, and not terrestrial   nature;
      
   {  {\bf 4}. ``non-correlation puzzle":}  Conventional picture for the UV diffuse radiation suggests that it must be correlated with the 100 $\mu m$ thermal  emission as both radiations are assumed to be related to the dust and its distribution in the galaxy. However, this conjecture dramatically fails  as the 100 $\mu m$ thermal  emission  is highly asymmetric and explicitly shows a strong correlation with localization of the  emitting stars, while UV diffuse emission is highly uniform and it is not correlated with the dust distribution, see Fig. 14  in \cite{Henry_2014}. 
  
  The authors of \cite{Henry_2014} conclude, I quote  ``the  source of the diffuse FUV emission is unknown --that is the mystery that is referred to in the title". I keep the same term ``mysterious radiation" in the title of the present work. Furthermore, the authors of \cite{Henry_2014}
  argued that this ``exotic" mysterious component of emission is unknown but must be galactic origin, and might be due to the interaction  of the DM with the nuclei of the interstellar medium. In addition, the authors   of \cite{Henry_2014}  also suggested that if the DM particle is electrically neutral but represents a composite system, than such DM particles  may contribute to the ``exotic"  diffuse UV component. In fact this proposal had been advocated even earlier in \cite{2012MmSAI..83..409H}. 
  
    In the present work we offer a specific DM model, the so-called AQN  dark matter  model which is capable to produce the required ``exotic"  diffuse UV radiation as mentioned in previous paragraph.   
      The AQN model  was  invented long ago    \cite{Zhitnitsky:2002qa} for drastically different purpose (unrelated   to the present studies)
 to   explain  the observed  similarity between the dark matter and the visible matter  densities in the Universe.
 The original motivation \cite{Zhitnitsky:2002qa} for the AQN  model can be explained in two lines as follows. 
It is commonly  assumed that the Universe 
began in a symmetric state with zero global baryonic charge 
and later 
(through some baryon number violating process, non- equilibrium dynamics, and $\cal{CP}$ violation effects, realizing three  famous   Sakharov's criteria \cite{Sakharov:1967dj}) 
evolved into a state with a net positive baryon number, which represents the  idea of baryogenesis.
In the AQN  framework the 
baryogenesis is actually a charge segregation  (rather than charge generation) process 
in which the global baryon number of the universe remains 
zero at all times.  
This scenario should be considered as  an 
alternative   path which is  qualitatively  distinct  from conventional baryogenesis.  The AQN construction in some respects is similar to the Witten's quark nuggets, see  \cite{Witten:1984rs,Farhi:1984qu,DeRujula:1984axn}. This type of DM  is ``cosmologically dark'' not because of the weakness of the  interactions with the visible  SM particles, but due to their small cross-section-to-mass ratio, which scales down many observable consequences of an otherwise strongly-interacting DM candidate.

The main distinct feature of the  AQN model which plays
absolutely crucial role for  the present work  is that nuggets can be made of {\it matter} as well as {\it antimatter} during the QCD transition
as a result of this charge segregation (rather than charge generation) process. 
In this scenario  the DM density, $\Omega_{\rm DM}$ representing the matter and anti-matter nuggets, and the visible    density, $\Omega_{\rm visible}$, will automatically assume the  same order of magnitude densities  $\Omega_{\rm DM}\sim \Omega_{\rm visible}$   as they are both proportional to one and the same fundamental dimensional parameter of the theory, the $\Lambda_{\rm QCD}$. Therefore, the AQN  model,   by construction, actually resolves two fundamental problems in cosmology    without necessary to fit any parameters of the model.  
\exclude{We refer to the original papers   \cite{Liang:2016tqc,Ge:2017ttc,Ge:2017idw,Ge:2019voa} devoted to the specific questions  related to the nugget's formation, generation of the baryon asymmetry, and 
survival   pattern of the nuggets during the evolution in  early Universe with its unfriendly environment. 
However,
} For the present studies  we  take the  agnostic viewpoint regarding  the questions on formation mechanism of the AQNs, and assume that such nuggets made of antimatter are present in our Universe, see   \cite{Zhitnitsky:2021iwg} for a short overview on specific questions  related to the nugget's formation, generation of the baryon asymmetry, and 
survival   pattern of the nuggets during the evolution in  early Universe with its unfriendly environment.   This assumption is consistent with all available constraints as long as  the average baryon charge of the nuggets is sufficiently large $\langle B\rangle \geq 3\cdot 10^{24}$. 

  \exclude{As we argue in the present work the features  listed as the puzzles $\bf 1-4$ can be naturally understood within this AQN framework. We refer to recent brief review  \cite{Zhitnitsky:2021iwg} on the AQN model for a general background,  while below we overview the basic  characteristic of the model which are crucial for the present work.
 }
 The presence of the {\it antimatter} nuggets in the system implies that there will be  annihilation events leading to  large number of observable effects on different scales: from galactic scales to the terrestrial rare events. In fact, there are many hints suggesting that such annihilation events may  indeed  took place in early Universe as well as they are happening now in present epoch. In particular, the  AQNs might be responsible for a resolution of   the  ``Primordial Lithium Puzzle" \cite{Flambaum:2018ohm}
 during Big Bang  Nucleosynthesis (BBN) epoch. The AQNs  may also  alleviate the tension between standard model cosmology and the recent EDGES observation of a stronger than anticipated 21 cm absorption feature as argued in \cite{Lawson:2018qkc}.   The AQNs might be also responsible for famous   long standing problem of  the ``Solar Corona Mystery"
  \cite{Zhitnitsky:2017rop,Raza:2018gpb} when the   so-called ``nanoflares" conjectured by Parker long ago \cite{Parker} are   identified with the  annihilation events in the AQN framework. The AQNs could be also responsible for mysterious and anomalous Cosmic Ray (CR) like events such as mysterious bursts \cite{Zhitnitsky:2020shd} observed by the Telescope Array collaboration or Multi Modal Clustering anomalous events  \cite{Zhitnitsky:2021qhj} observed by the HORIZON 10T collaboration.  Sufficiently large (and vary rare) AQNs with $B>10^{27}$ entering  the Earth's atmosphere could produce  infrasound and seismic acoustic waves     as discussed in   \cite{Budker:2020mqk}  when the infrasound and seismic acoustic waves indeed have been recorded  by  dedicated instruments while a synchronized  all-sky camera network (visible frequency bands)    ruled out a conventional meteor source.

The list   with numerous  deviations from the standard cosmology  at different scales     which could be explained in terms of  the same  antimatter nuggets within the same framework as mentioned   above  is already quite long, but definitely  is far from being  complete. 
Our comment here  is that the presence of the antimatter nuggets directly impacts all these (seemingly unrelated) fields of physics,  from baryogenesis (replaced by charge segregation)  and the  ``Primordial Lithium Puzzle" in early Universe  to excess in diffuse FUV background and rare CR-like anomalous events at present time. If the future observations and more precise  data confirm  these hints on deviations from the standard astrophysics and cosmology it will  be  an extraordinary evidence  revealing  the nature of the dark matter.

\section{ The AQN   Dark Matter model}  
The typical   baryon number density of such dense objects is the same order of magnitude as the conventional nuclear matter density $n\sim 10^{40}\rm cm^{-3}$, while typical   baryon charge of the nugget is around $B\sim 10^{25}$ which implies that the mass of a typical nugget is around $M\sim10 $ g, while its size $R\sim 10^{-5}$ cm, see below with more specific and technical  details.    

 If  the AQNs  enter the regions of the surrounding material (such as stars, planets or interstellar medium) the  annihilation processes start and the internal temperature of the nuggets $T$ starts to rise. A typical internal temperature   of  the  AQNs for the  galactic environment
of density $n$  can be estimated from the condition that
 the radiative output, the surface emissivity   $F_{\text{tot}}(T)$  must balance the flux of energy  due to the annihilation processes  \cite{Forbes:2008uf}, i.e.
 \be
\label{eq:rad_balance}
   4\pi R^2 \cdot F_{\text{tot}} (T)
\approx \kappa\cdot  (\pi R^2) \cdot (2~ {\rm GeV})\cdot n \cdot v_{\rm AQN},  
\ee 
where the left hand side accounts for the total energy radiation from the  AQN's surface per unit time  while  
 the right hand side  accounts for the rate of annihilation events when each successful annihilation event of a single baryon charge produces $\sim 2m_pc^2\approx 2~{\rm GeV}$ energy. In Eq.\,(\ref{eq:rad_balance}) we assume that  the nugget is characterized by the geometrical cross section $\pi R^2$ when it propagates 
in environment with local density $n$ with velocity $v_{\rm AQN}\sim 10^{-3}c$.  The factor $\kappa$ is introduced to account for the fact that not all matter striking the  AQN will 
annihilate and not all of the energy released by an annihilation will be thermalized in the  AQNs by changing the internal temperature $T$. 
In particular,  some portion of the energy will be released in form of the axions, neutrinos, and also in form of the x-rays from the so-called hot spots, see details below. 
The high probability 
of reflection at the sharp quark matter surface lowers the value of $\kappa$. The ionization of the AQNs, on the other hand,    may induce the negative charge of the AQNs, and as a consequence may increase the value of $\kappa$ by  effectively increasing the   cross section (\ref{eq:rad_balance}). Furthermore, the density of the galactic environment $n ({\bf{r}})$ changes dramatically from  region to region, which implies that the temperature of the nuggets also varies  correspondingly, which also introduces a large uncertainty.   

One can dramatically reduce  all these numerous uncertainties  in the  estimation for  the effective temperature $T_{\rm eff}$   by normalizing the intensity of the emission to some  specific observations which can be identified with the AQN annihilation events in interstellar medium. In fact, it has been argued in \cite{Forbes:2006ba} that the diffuse x ray  emission analyzed in \cite{Muno:2004bs}   cannot be explained in terms of  any conventional astrophysical sources  but can be naturally explained in terms of  the AQN annihilation events. The combination of the arguments  from \cite{Forbes:2008uf,Forbes:2006ba} then suggest that the  effective temperature of the AQNs in the central region of the galaxy could be very high $T_{\rm eff} \geq (2.5-5) ~\rm eV$, see estimate below. Such high temperature for  $T_{\rm eff}$  implies that these nuggets   could be   very strong UV emitters, which is precisely the topic of the present work.  

To  proceed with our estimates  we need 
  the expression for the spectral surface emissivity due to the bremsstrahlung radiation from electrosphere at temperature $T$.   It  has been computed in \cite{Forbes:2008uf}:
\be 
  \label{eq:P}
  \frac{d{F}}{d{\omega}}(\omega) \approx
  \frac{4}{45}
  \frac{T^3\alpha^{5/2}}{\pi}\sqrt[4]{\frac{T}{m}}\cdot 
 f(x), ~~~~ x\equiv \frac{\omega}{T}
\ee
where dimensionless function $f(x)$ with sufficient accuracy can be  approximated as follows:
\begin{equation}
  f(x) \approx \begin{cases}
     \left(1+x\right)e^{-x} \left(17-12\ln(x/2)\right) & x<1,\\
     \left(1+x\right)e^{-x}\left(17+12\ln(2)\right) & x\geq1.
    \nonumber
  \end{cases}
\end{equation}
The total surface
emissivity entering (\ref{eq:rad_balance}) is given by:
\begin{equation}
  \label{eq:P_t}
  F_{\text{tot}}(T) = 
     \int^{\infty}_0\!\!\!\!\!d{\omega}\;
  \frac{d{F}}{d{\omega}}(\omega) 
  \approx
  \frac{16}{3}
  \frac{T^4\alpha^{5/2}}{\pi}\sqrt[4]{\frac{T}{m}}.\\
\end{equation}
The most important feature of this bremsstrahlung radiation is that, in contrast with black body radiation,  it is very flat for $\omega\leq T$, while it starts   to diminish only for sufficiently high frequencies $\omega \gg T$.  Numerically, the intensity of radiation (\ref{eq:P_t}) is also much smaller than black body radiation by factor $\sim 10^{-6}$ for relevant temperatures around $T\sim $ eV.

To compute the observable flux $ \Phi_{\Gamma} $ for   a given   frequency band from all nuggets one should multiply (\ref{eq:rad_balance}) to the number density of the nuggets $n_{\rm DM}\simeq \rho_{\rm DM}/(m_pB)$ and integrate 
along the line of sight $\Gamma$, i.e.
\begin{equation}
  \label{eq:direct_integral}
  \Phi_{\Gamma}  \propto 
  \int_{\Gamma}   d{l}\, n (l)n_{\rm DM}(l),
\end{equation}
where we assume that all other parameters, except density $n$,  entering the right hand side of (\ref{eq:rad_balance})
are approximately the same.  The key point here is that the 
 comparison between observations for different frequency bands   along the same line-of-sight
is possible because the local emission  for  any frequency band depends only on the
local rate of annihilation $ \propto
n ({\bf{r}})n_{\rm DM}(\bf{r})$ as the spectrum is fixed by (\ref{eq:P}).  Therefore, the main uncertainties (mentioned above 
and related to the AQN dynamics as well as to the matter and DM distributions)   will be cancelled  out
when comparing the emissions from the same position in the
sky.

\section{ The FUV  diffuse emission} 
From the arguments outlined above one can represent the spectral emissivity with frequency $2\pi\nu=x T$ produced by AQNs at arbitrary direction in sky determined by path $\Gamma$ as follows
\be
\label{emissivity}
\frac{d \Phi_{\Gamma} (x)}{dx}\approx \bar{\Phi}\left[\frac{f(x)}{\int_0^{\infty} f(x)dx} \right]\left[\frac{\int_{\Gamma} d{l}\, n (l)n_{\rm DM}(l)}{ \int_{\bar{\Gamma}}   d{\bar{l}}\, n (\bar{l})n_{\rm DM}(\bar{l})} \right] \left[\frac{1-g}{g} \right].~~~
\ee
In expression (\ref{emissivity}) the parameter $g$ is unknown empirical factor of order of one to be discussed below while the main normalization factor $\bar{\Phi}$ is chosen as the  excess of the diffused X-ray emission from the galactic core as observed by Chandra and analyzed in \cite{Muno:2004bs}.  This choice is based on the following arguments \cite{Forbes:2006ba}.   
The X-ray emissions    from the galactic core
provide a puzzling picture: they seem to indicate that an 8 keV thermal plasma is being maintained, but the source of energy fuelling
this plasma is a mystery.  After subtracting known X-ray sources from
the Chandra  X-ray images of the galactic core, one finds a residual
diffuse thermal X-ray emission with a thermal component well described
by a hot 8 keV  plasma with surface brightness $\Phi_{\rm keV} = (1.8$ --
$3.1)\times 10^{-6}$~erg/cm$^2$/s/sr~\cite{Muno:2004bs}.  To sustain
this plasma, some $10^{40}$~erg/s of energy must be supplied to the
galactic core which is much more than the observed rate of supernovae,
for example, can explain~\cite{Muno:2004bs}. 

In    has been  argued in \cite{Forbes:2006ba} that the spectrum and the intensity of the AQN induced  X-ray emission 
  fits the observations, which is the crucial argument  for our normalization\footnote{Another way to normalize the intensity (\ref{emissivity}) is to use the observed excess in 511 keV line  as it was done in the original paper  \cite{Forbes:2006ba}. However, in this case an extra uncertainty appears due to unknown relation between 511 keV line emission  which is due to $e^+e^-$ annihilation  and the baryon charge annihilation processes within AQN core. In fact it has been recently claimed \cite{Flambaum:2021xub} that the mechanism for the $e^+e^-$ annihilation is different from the original proposal \cite{Oaknin:2004mn,Zhitnitsky:2006tu} as the annihilation
of the visible electrons with AQN's positrons may occur far away from the nugget's location due to the strong ionization features of the AQNs. Alternatively, one could normalize (\ref{emissivity}) in terms of DM and visible matter distributions as it was done in \cite{Oaknin:2004mn}. However,   it would bring an  enormous uncertainty in normalization, which we try to avoid by normalizing to the observed flux.}   in terms of factor $\bar{\Phi}$ entering (\ref{emissivity}):   
  \begin{equation}
  \label{observation}
   \bar{\Phi} \approx (2.5-3.9)\cdot
    10^{-5}\frac{\text{erg}}{\text{cm}^2\cdot\text{s}},~~~~ [\rm   X-ray~observations]
  \end{equation}
  where we multiply by factor $4\pi$ the observed intensity $\Phi_{\rm keV}$ extracted in \cite{Muno:2004bs} as quoted above. 
  As discussed in  \cite{Forbes:2006ba} the X-ray emission in the AQN framework is originated from the so-called ``hot spots" 
  which represent the very localized regions close to the nugget's surface, where the annihilation events  take place. The corresponding portion of the energy released from ``hot spots" is parameterized by unknown factor $g\ll 1 $ entering (\ref{emissivity}). The dominant portion  of the energy $(1-g)$ due to the annihilation will be thermalized, and will be emitted with the spectrum determined by  (\ref{eq:P}). It explains the numerical factor $(1-g)/g$ entering 
  (\ref{emissivity}). The last factor $\bar{\Gamma}$  
  which requires the explanation represents a specific path $\bar{\Gamma}$ in the direction  to the galactic core where uniform X-ray diffused flux  (\ref{observation}) is recorded. We also use   notations $\bar{l}$ for  the integration variable for the path $\bar{\Gamma}$ 
 to be distinguished   from   generic path $\Gamma$.
  
  Now we want to argue that a typical temperature $T_{\rm eff}$ which implicitly (through parameter $x\equiv \omega/T_{\rm eff}$) enters (\ref{emissivity}) indeed typically assumes the values in  the eV range such that the AQNs will be the source of  the diffuse UV radiation, which represents the main claim of this work. 
  Indeed, to avoid a large number of  uncertainties mentioned above we can estimate $T_{\rm eff}$ by comparing the excess of radiation observed by   WMAP  collaboration and coined as the 
    WMAP  haze  \cite{Finkbeiner:2003im,Finkbeiner:2004je,PhysRevD.76.083012,Hooper:2007gi,Dobler:2007wv} and the X-ray diffused radiation we already mentioned. Assuming that  the observed  WMAP  haze intensity is  saturated\footnote{It has been argued in \cite{Finkbeiner:2004je} that  the    WMAP  haze can be attributed to the spinning dust or the dark matter annihilation \cite{PhysRevD.76.083012,Hooper:2007gi}. As the physical source of this excess of GHz emission  remains to be  a matter of debate  we assume in our estimates (\ref{eq:chandra-wmap}) that a finite fraction of this excess is due to the AQN annihilation events as argued in  \cite{Forbes:2008uf}.} by
the  AQN  radiation  as advocated in \cite{Forbes:2008uf} with the flat spectrum (\ref{eq:P}) extending from UV to GHz emission, one arrives to the following estimate for $T_{\text{eff}}$ (quoted  above) in terms of the   parameter $g$:
\begin{equation}
  \label{eq:chandra-wmap}
  \frac{\text{eV}}{T_{\text{eff}}}\cdot\frac{1-g}{g} \approx
  (2 \text{--} 4) ~~~ \Rightarrow ~~ T_{\text{eff}}\geq (2.5-5) \rm ~eV,
\end{equation}
where for numerical estimates we use $g\simeq 0.1$.  One should emphasize that the uncertainty  of parameter $g$ 
is not a deficiency of the AQN framework itself. Rather, this inevitable uncertainty results from very complex annihilation  pattern of the baryons from visible 
material (confined hadronic phase) with antimatter from the AQNs (colour superconducting phase). Therefore, this parameter $g$ fundamentally cannot be computed from the first principles and will be treated in what follows as empirical phenomenological parameter.

The key point here is that both emissions,  the     WMAP  haze as well as diffuse X-ray emissions are originated from approximately the same region in the sky, and both are related  to the same annihilation processes of the baryon charge   within the AQN framework such that the  only uncertainty in the estimate of the $T_{\text{eff}}$ comes   from parameter $g$ defined above\footnote{Of course  a number of  uncertainties related to extracting   WMAP  haze intensity and the X ray intensity remain. Our claim refers to the theoretical uncertainties related to the AQN model itself and its interaction with environment.}. 

It is very instructive to compare our estimates determined by  (\ref{emissivity}) and  (\ref{observation}) with intensities  of the diffuse FUV  emission as recorded by \cite{Henry_2014,Akshaya_2018,2019MNRAS.489.1120A}.   To proceed with such comparison we have to use the same  units.  The simplest way to compare is to use  the conversion worked out in  \cite{Henry_2014}
where, I quote ``the intensity of excess FUV radiation detected by GALEX over its bandpass (1380-2500)$\rm \AA$ is 
 $ 10^{-5} {\text{erg}}~{\text{cm}^{-2}~\text{s}^{-1}}$, assuming a flat spectrum with 300 photons $  {\rm ({cm}^{-2}~ {s}^{-1} sr^{-1}\AA^{-1})}$."
     
     This implies that the magnitude of 300 photons   which typically emerges  in analysis \cite{Henry_2014,Akshaya_2018,2019MNRAS.489.1120A} corresponds to the intensity  $ 10^{-5} {\text{erg}}~{\text{cm}^{-2}~\text{s}^{-1}}$ which appears in our   prediction  (\ref{emissivity})  with normalization  factor  $\bar{\Phi}$ defined by (\ref{observation}). 
     Several  numerical factors   modify this crude       estimate.  
 First, there is an intensity suppression   due to decreasing of the interstellar medium density $n(l)$   entering  (\ref{emissivity}) for  the path 
 $\Gamma$ for FUV (where the excess with 300 photons is measured) in comparison with  path $\bar{\Gamma}$. This suppression factor might be largely  neutralized  
   by the  enhancement factor $1/g\sim 10$ entering  (\ref{emissivity}). The same factor $g^{-1}$ may also increase the effective temperature $T_{\text{eff}}$  comparison with oversimplified estimate  (\ref{eq:chandra-wmap}).   Furthermore,   the $T_{\text{eff}}$ might assume somewhat  higher values in the galactic core regions,  which also contribute to FUV measured at higher latitudes due to the re-scattering from the dust.  
   Important point here is that the observed   FUV intensity   is consistent with the temperature range  (\ref{eq:chandra-wmap})
   as the spectrum given by $f(x)$  is very flat and effectively starts to diminish only  for sufficiently high frequencies $2\pi \nu \geq T_{\text{eff}}$.
   
   Precise portion of the emission in the FUV   frequency band within the window  $\Delta \nu$ is determined by the parameter $\chi$ defined as follows:
  \be
  \label{chi}
  \chi\equiv \frac{\int_{x_{\rm  min}}^{ x_{\rm max}} dx f(x)}{\int_0^{\infty}dxf(x)}, ~~~  \left(\frac{2\pi \nu}{T_{\rm eff}}\right)\in(x_{\rm  min}, x_{\rm max}).
  \ee 
   The suppression factor $\chi$ obviously dramatically depends on effective internal temperature $T_{\rm eff}$ of the nuggets, which itself is determined by the local    interstellar medium density $n$ according to (\ref{eq:rad_balance}). As an  illustration let us consider   a typical  temperature $T_{\text{eff}}\simeq 5$ eV as estimated in (\ref{eq:chandra-wmap}) and bandpass (1380-2500)$\rm \AA$ where 
 the excess of FUV radiation was detected by GALEX.   The factor $\chi$ in this case   is about  $\chi\simeq 0.2$, which  implies that  quite substantial   portion   of the energy will be emitted by the AQNs in this relatively narrow frequency band.

  Our estimates  presented above suggest    that
  the AQN-induced FUV radiation may naturally assume $ \sim 10^{-5} {\text{erg}}~{\text{cm}^{-2}~\text{s}^{-1}}$ intensity range. However, those estimates were based on  comparison with  emission in  a different frequency band (X-ray emission) which is assumed to be also the AQN-induced radiation. Now we want to support the basic normalization factor (\ref{emissivity}) by  independent direct estimates  without referring to any other  observed diffuse  radiation.  
  
     First, we estimate a total luminosity $ L_{\rm AQN}$ from a single AQN by multiplying (\ref{eq:P_t}) to the surface area:
     \be
     \label{AQN_luminosity}
    L_{\rm AQN}\approx  4\pi R^2 \cdot F_{\text{tot}} (T)\approx 4.7\left(\frac{T}{5~ \rm eV}\right)^{\frac{17}{4}}\rm \frac{erg}{s},
     \ee
     where for numerical estimates we use $R\approx 2.2\cdot 10^{-5} \rm cm$ and  $B\approx 10^{25}$ as in   our previous studies
   for  other applications in this model, see \cite{Zhitnitsky:2021iwg} for review. 
     To estimate the total FUV intensity  $ \Phi^{FUV}_{\rm AQN}$ produced by all AQNs one should multiply the luminosity $ L_{\rm AQN}$  to the nugget's density $n_{\rm AQN}\sim \rho_{\rm DM}/(m_pB)$ and the mean free path ${\cal{R}}$ in the local interstellar medium, i.e. 
          \be
     \label{FUV-AQN}
     \Phi^{FUV}_{\rm AQN}\sim  L_{\rm AQN}  n_{\rm AQN}  {\cal{R}}  \chi\sim 5\cdot 10^{-5}\left(\frac{T}{5~ \rm eV}\right)^{\frac{17}{4}} \rm \frac{erg}{s\cdot  cm^2},~~~~
     \ee
     where  for numerical estimates we take the conventional value for the DM density $\rm \rho_{\rm DM}\simeq 0.3 ~GeV/cm^3$. We also  used parameter   $\chi\simeq 0.2$ as estimated above. This relatively large value for $\chi$ 
     is a result of the  ``flatness"  of the bremsstrahlung radiation as mentioned after (\ref{eq:P_t}).    We also adopted   parameter ${\cal{R}}\simeq 0.6 \rm~ kpc$  used in  \cite{Henry_2014}.   It  represents the result of   conventional   estimates for the  mean free path of the FUV photons propagating in the galaxy.

    It is instructive to compare the estimate (\ref{FUV-AQN}) with    analogous  estimates    for WIMP- like DM particles  performed  in \cite{Henry_2014}.  The authors  of  \cite{Henry_2014} claim that the total  FUV intensity produced by WIMPs cannot exceed $10^{-17} \rm  {erg}/{(s\cdot cm^2)}$  which according to the authors of  \cite{Henry_2014}  represents a   model-independent upper limit  for WIMP-like  models.
     It should be  contrasted with  our AQN based estimates  which show a   perfect consistency  with the observed intensity on the level $10^{-5} \rm  {erg}/{(s\cdot cm^2)}$ as well as  with estimates (\ref{emissivity}), (\ref{observation}) based on normalization to    the X ray diffuse emission. 

We are now in position to explain how the puzzles {\bf 1-4} listed above are naturally resolved  within AQN framework. 
     First of all the ``uniformity puzzle" is obviously resolved as the DM in form of the AQNs are uniformly distributed in the galaxy. 
     Similarly, the ``galactic longitude puzzle" is resolved as the DM particles are not sensitive to the galactic longitude, as the AQNs  are not correlated with  locations of the UV emitting stars.    The ``galactic latitude puzzle" is also naturally resolved as the  increase  in brightness toward lower     galactic latitude is a direct consequence of the AQN framework when the number of annihilation events is proportional to 
interstellar  matter density according to (\ref{eq:direct_integral}) which obviously leads to the increase  of the temperature $T$ and as the consequence to the brightness toward lower     galactic latitude.  Finally, 
the ``non-correlation puzzle" with the 100 $\mu m$ thermal  emission is resolved as the dominant fraction of the 100 $\mu m$ radiation follows the locations  of the  emitting stars, in contrast  with AQNs, which contribute very little to the thermal 100 $\mu m$ radiation\footnote{Indeed, a typical observed value for the 100 $\mu m$ emission line  ranges in $(0.7-2.5) \rm MJy/sr$  depending on location, see Figs. 12, 13 in \cite{Henry_2014}. This observed intensity is numerically much larger  than the AQN induced contribution to this frequency band as determined by the spectrum  (\ref{eq:P}) with normalization (\ref{emissivity}).}.
One should emphasize that a conventional WIMP-type DM models are not capable to generate the required intensity of the signal as it was already pointed out in \cite{Henry_2014}, as it was already mentioned. 
 The AQNs, in contrast with WIMP-type  particles, are   the macroscopically large   objects made of strongly interacting quarks and gluons of the Standard Model (SM). In this sense it is  indeed a ``composite system" in terminology  of \cite{Henry_2014}.  
  
\section {Concluding comments} 
There are several profound  consequences   of our proposal identifying   the mysterious diffuse UV   radiation \cite{Henry_2014,Akshaya_2018,2019MNRAS.489.1120A}     with the emissions by hot AQNs.
First of all this proposal may also shed some light on  a long standing problem  related to  the  nature   of the re-ionization of the   Universe. Indeed,  by studying the variety of spectra the authors of   \cite{Henry:2018yar} argued  that the same DM source which is responsible for FUV diffuse emission 
(which is the topic of  this work) must also emit, I quote:  ``a continuum of photons in the range $\sim 850 \rm \AA$ to about 2000 $\rm \AA$". This feature is obviously present in the AQN framework as the spectrum (\ref{eq:P}) is very broad and for $T_{\rm eff}\geq 5$ eV extends to the shorter wavelengths well beyond  the cut off at the $  912 \rm \AA$ when the hydrogen atom can be ionized as this wavelength exactly corresponds to the Rydberg constant $13.6$ eV. In different words, the same AQNs could be the source of ionizing radiation which is  known to be present  well above the galactic plane \cite{Henry:2018yar}. 

Furthermore, the same DM source in form of the AQNs may also contribute to the resolution of  another long standing problem 
related to  the Extragalactic Background Light (EBL). Indeed, it has been known for some time that the conventional measurements    cannot be explained by diffuse galaxy halos or intergalactic stars. The discrepancy could be as large as  factor $\sim (2-3)$ or even more, see e.g. recent review \cite{Mattila:2019ybk}. Our comment here is that the AQNs may fulfill this shortage as the spectrum   (\ref{eq:P}) is very broad and  includes optical and IR light.

 There is a number of direct tests which can be performed in  future   to substantiate  or refute this proposal identifying   the mysterious diffuse UV   radiation \cite{Henry_2014,Akshaya_2018,2019MNRAS.489.1120A}     with the emission by AQNs.   Some specific  suggestions  for future studies have been already    discussed in  \cite{Henry_2014,Akshaya_2018,2019MNRAS.489.1120A,Henry:2018yar}, and   include such instruments as the Alice UV spectrometer aboard the New Horizon mission, and we have nothing new to add here. The only original comment we would like to make is  that the intensity and spectral  features of the radiation  in the AQN framework are determined by the line of sight  which includes  both:
the DM and visible matter distributions according to (\ref{eq:direct_integral}). It should be   contrasted with conventional WIMP-like models when it is exclusively determined by the DM distribution.  Therefore, some specific morphological correlations with DM and visible matter distributions can be explicitly studied in future.

It is quite amazing that the spectral features and intensities for ``exotic" sources which were required (anticipated)  by analysis \cite{Henry_2014,Akshaya_2018,2019MNRAS.489.1120A,Henry:2018yar}  in form of the ``composite system" in terminology \cite{Henry_2014} to fit    the observations are automatically present in the AQN framework which was originally invented for dramatically different purposes with very  different motivation, see \cite{Zhitnitsky:2021iwg} for review. 
One should also emphasize that   in the present  work we use the same basic parameters  which were previously used   for    very different  applications in drastically  different environment without any attempt to modify these parameters  to better fit the observations. We consider this ``miracle coincidence" as a strong argument supporting this proposal.

    If  our  interpretation on   source of the excess of the diffuse FUV emission  is confirmed by future studies
 it would represent  an extraordinary       evidence supporting the resolution of   two long standing puzzles:   it reveals  the nature of the DM   and the   matter-antimatter asymmetry of  our Universe as these two problems of the cosmology   are intimately  linked in the AQN framework.

    \section*{Acknowledgements}
 This research was supported in part by the Natural Sciences and Engineering
Research Council of Canada.

\bibliography{FUV}

\begin{thebibliography}{10}
\expandafter\ifx\csname url\endcsname\relax
  \def\url#1{\texttt{#1}}\fi
\expandafter\ifx\csname urlprefix\endcsname\relax\def\urlprefix{URL }\fi
\expandafter\ifx\csname href\endcsname\relax
  \def\href#1#2{#2} \def\path#1{#1}\fi

\bibitem{Henry_2014}
R.~C. Henry, J.~Murthy, J.~Overduin, J.~Tyler,
  \href{https://doi.org/10.1088/0004-637x/798/1/14}{{The} {Mystery} {of} {the}
  {cosmic} {diffuse} {ultraviolet} {background} {radiation}}, \apj 798~(1)
  (2014) 14.
\newblock \href {http://dx.doi.org/10.1088/0004-637x/798/1/14}
  {\path{doi:10.1088/0004-637x/798/1/14}}.
\newline\urlprefix\url{https://doi.org/10.1088/0004-637x/798/1/14}

\bibitem{Akshaya_2018}
M.~S. Akshaya, J.~Murthy, S.~Ravichandran, R.~C. Henry, J.~Overduin,
  \href{https://doi.org/10.3847/1538-4357/aabcb9}{The diffuse radiation field
  at high galactic latitudes}, \apj 858~(2) (2018) 101.
\newblock \href {http://dx.doi.org/10.3847/1538-4357/aabcb9}
  {\path{doi:10.3847/1538-4357/aabcb9}}.
\newline\urlprefix\url{https://doi.org/10.3847/1538-4357/aabcb9}

\bibitem{2019MNRAS.489.1120A}
M.~S. {Akshaya}, J.~{Murthy}, S.~{Ravichandran}, R.~C. {Henry}, J.~{Overduin},
  {Components of the diffuse ultraviolet radiation at high latitudes}, Mon.\
  Not.\ R.\ Astron.\ Soc. 489~(1) (2019) 1120--1126.
\newblock \href {http://arxiv.org/abs/1908.02260} {\path{arXiv:1908.02260}},
  \href {http://dx.doi.org/10.1093/mnras/stz2186}
  {\path{doi:10.1093/mnras/stz2186}}.

\bibitem{Zhitnitsky:2002qa}
A.~R. {Zhitnitsky}, {`Nonbaryonic' dark matter as baryonic colour
  superconductor}, \jcap 10 (2003) 010.
\newblock \href {http://arxiv.org/abs/hep-ph/0202161}
  {\path{arXiv:hep-ph/0202161}}, \href
  {http://dx.doi.org/10.1088/1475-7516/2003/10/010}
  {\path{doi:10.1088/1475-7516/2003/10/010}}.

\bibitem{2012MmSAI..83..409H}
R.~C. {Henry}, {Progress in understanding the diffuse UV cosmic background},
  Mem.S.A.It. 83 (2012) 409.
\newblock \href {http://arxiv.org/abs/1205.0430} {\path{arXiv:1205.0430}}.

\bibitem{Sakharov:1967dj}
A.~Sakharov, {Violation of CP Invariance, C asymmetry, and baryon asymmetry of
  the universe}, JETP Lett. 5 (1967) 24--27.
\newblock \href {http://dx.doi.org/10.1070/PU1991v034n05ABEH002497}
  {\path{doi:10.1070/PU1991v034n05ABEH002497}}.

\bibitem{Witten:1984rs}
E.~{Witten}, {Cosmic separation of phases}, \prd 30 (1984) 272--285.
\newblock \href {http://dx.doi.org/10.1103/PhysRevD.30.272}
  {\path{doi:10.1103/PhysRevD.30.272}}.

\bibitem{Farhi:1984qu}
E.~{Farhi}, R.~L. {Jaffe}, {Strange matter}, \prd 30 (1984) 2379--2390.
\newblock \href {http://dx.doi.org/10.1103/PhysRevD.30.2379}
  {\path{doi:10.1103/PhysRevD.30.2379}}.

\bibitem{DeRujula:1984axn}
A.~{De Rujula}, S.~L. {Glashow}, {Nuclearites - A novel form of cosmic
  radiation}, \nat 312 (1984) 734--737.
\newblock \href {http://dx.doi.org/10.1038/312734a0}
  {\path{doi:10.1038/312734a0}}.

\bibitem{Zhitnitsky:2021iwg}
A.~Zhitnitsky, {Axion quark nuggets. Dark matter and
  matter\textendash{}antimatter asymmetry: Theory, observations and future
  experiments}, Mod. Phys. Lett. A 36~(18) (2021) 2130017.
\newblock \href {http://arxiv.org/abs/2105.08719} {\path{arXiv:2105.08719}},
  \href {http://dx.doi.org/10.1142/S0217732321300172}
  {\path{doi:10.1142/S0217732321300172}}.

\bibitem{Flambaum:2018ohm}
V.~V. Flambaum, A.~R. Zhitnitsky, {Primordial Lithium Puzzle and the Axion
  Quark Nugget Dark Matter Model}, Phys. Rev. D 99~(2) (2019) 023517.
\newblock \href {http://arxiv.org/abs/1811.01965} {\path{arXiv:1811.01965}},
  \href {http://dx.doi.org/10.1103/PhysRevD.99.023517}
  {\path{doi:10.1103/PhysRevD.99.023517}}.

\bibitem{Lawson:2018qkc}
K.~Lawson, A.~R. Zhitnitsky, {The 21 cm absorption line and the axion quark
  nugget dark matter model}, Phys. Dark Univ. 24 (2019) 100295.
\newblock \href {http://arxiv.org/abs/1804.07340} {\path{arXiv:1804.07340}},
  \href {http://dx.doi.org/10.1016/j.dark.2019.100295}
  {\path{doi:10.1016/j.dark.2019.100295}}.

\bibitem{Zhitnitsky:2017rop}
A.~{Zhitnitsky}, {Solar Extreme UV radiation and quark nugget dark matter
  model}, \jcap 10 (2017) 050.
\newblock \href {http://arxiv.org/abs/1707.03400} {\path{arXiv:1707.03400}},
  \href {http://dx.doi.org/10.1088/1475-7516/2017/10/050}
  {\path{doi:10.1088/1475-7516/2017/10/050}}.

\bibitem{Raza:2018gpb}
N.~Raza, L.~van Waerbeke, A.~Zhitnitsky, {Solar corona heating by axion quark
  nugget dark matter}, Phys. Rev. D 98~(10) (2018) 103527.
\newblock \href {http://arxiv.org/abs/1805.01897} {\path{arXiv:1805.01897}},
  \href {http://dx.doi.org/10.1103/PhysRevD.98.103527}
  {\path{doi:10.1103/PhysRevD.98.103527}}.

\bibitem{Parker}
E.~N. {Parker}, {Nanoflares and the solar X-ray corona}, \apj 330 (1988)
  474--479.
\newblock \href {http://dx.doi.org/10.1086/166485} {\path{doi:10.1086/166485}}.

\bibitem{Zhitnitsky:2020shd}
A.~Zhitnitsky, {The Mysterious Bursts observed by Telescope Array and Axion
  Quark Nuggets}, J.Phys.G:Nucl.Part.Phys.\href
  {http://arxiv.org/abs/2008.04325} {\path{arXiv:2008.04325}}, \href
  {http://dx.doi.org/10.1088/1361-6471/abd457}
  {\path{doi:10.1088/1361-6471/abd457}}.

\bibitem{Zhitnitsky:2021qhj}
A.~Zhitnitsky, {Multi-Modal Clustering Events observed by Horizon-10T and Axion
  Quark Nuggets}, Universe 7 (2021) 384.
\newblock \href {http://arxiv.org/abs/2108.04826} {\path{arXiv:2108.04826}},
  \href {http://dx.doi.org/10.3390/universe7100384}
  {\path{doi:10.3390/universe7100384}}.

\bibitem{Budker:2020mqk}
D.~Budker, V.~V. Flambaum, A.~Zhitnitsky, {Infrasonic, acoustic and seismic
  waves produced by the Axion Quark Nuggets}, Symmetry 14 (2022) 459.
\newblock \href {http://arxiv.org/abs/2003.07363} {\path{arXiv:2003.07363}},
  \href {http://dx.doi.org/10.3390/sym14030459}
  {\path{doi:10.3390/sym14030459}}.

\bibitem{Forbes:2008uf}
M.~M. {Forbes}, A.~R. {Zhitnitsky}, {WMAP haze: Directly observing dark
  matter?}, \prd 78~(8) (2008) 083505.
\newblock \href {http://arxiv.org/abs/0802.3830} {\path{arXiv:0802.3830}},
  \href {http://dx.doi.org/10.1103/PhysRevD.78.083505}
  {\path{doi:10.1103/PhysRevD.78.083505}}.

\bibitem{Forbes:2006ba}
M.~M. Forbes, A.~R. Zhitnitsky, {Diffuse x-rays: Directly observing dark
  matter?}, JCAP 01 (2008) 023.
\newblock \href {http://arxiv.org/abs/astro-ph/0611506}
  {\path{arXiv:astro-ph/0611506}}, \href
  {http://dx.doi.org/10.1088/1475-7516/2008/01/023}
  {\path{doi:10.1088/1475-7516/2008/01/023}}.

\bibitem{Muno:2004bs}
M.~P. Muno, F.~K. Baganoff, M.~W. Bautz, E.~D. Feigelson, G.~P. Garmire, M.~R.
  Morris, S.~Park, G.~R. Ricker, L.~K. Townsley, {Diffuse x-ray emission in a
  deep Chandra image of the Galactic center}, Astrophys. J. 613 (2004)
  326--342.
\newblock \href {http://arxiv.org/abs/astro-ph/0402087}
  {\path{arXiv:astro-ph/0402087}}, \href {http://dx.doi.org/10.1086/422865}
  {\path{doi:10.1086/422865}}.

\bibitem{Flambaum:2021xub}
V.~V. Flambaum, I.~B. Samsonov, {Radiation from matter-antimatter annihilation
  in the quark nugget model of dark matter}, Phys. Rev. D 104~(6) (2021)
  063042.
\newblock \href {http://arxiv.org/abs/2108.00652} {\path{arXiv:2108.00652}},
  \href {http://dx.doi.org/10.1103/PhysRevD.104.063042}
  {\path{doi:10.1103/PhysRevD.104.063042}}.

\bibitem{Oaknin:2004mn}
D.~H. {Oaknin}, A.~R. {Zhitnitsky}, {511 keV Photons from Color Superconducting
  Dark Matter}, \prl 94~(10) (2005) 101301.
\newblock \href {http://arxiv.org/abs/hep-ph/0406146}
  {\path{arXiv:hep-ph/0406146}}, \href
  {http://dx.doi.org/10.1103/PhysRevLett.94.101301}
  {\path{doi:10.1103/PhysRevLett.94.101301}}.

\bibitem{Zhitnitsky:2006tu}
A.~{Zhitnitsky}, {Width of the 511 keV line from the bulge of the galaxy}, \prd
  76~(10) (2007) 103518.
\newblock \href {http://arxiv.org/abs/astro-ph/0607361}
  {\path{arXiv:astro-ph/0607361}}, \href
  {http://dx.doi.org/10.1103/PhysRevD.76.103518}
  {\path{doi:10.1103/PhysRevD.76.103518}}.

\bibitem{Finkbeiner:2003im}
D.~P. Finkbeiner, {Microwave ism emission observed by wmap}, Astrophys. J. 614
  (2004) 186--193.
\newblock \href {http://arxiv.org/abs/astro-ph/0311547}
  {\path{arXiv:astro-ph/0311547}}, \href {http://dx.doi.org/10.1086/423482}
  {\path{doi:10.1086/423482}}.

\bibitem{Finkbeiner:2004je}
D.~P. Finkbeiner, G.~I. Langston, A.~H. Minter, {Microwave ISM emission in the
  Green Bank Galactic Plane Survey: Evidence for spinning dust}, Astrophys. J.
  617 (2004) 350--359.
\newblock \href {http://arxiv.org/abs/astro-ph/0408292}
  {\path{arXiv:astro-ph/0408292}}, \href {http://dx.doi.org/10.1086/425165}
  {\path{doi:10.1086/425165}}.

\bibitem{PhysRevD.76.083012}
D.~Hooper, D.~P. Finkbeiner, G.~Dobler,
  \href{https://link.aps.org/doi/10.1103/PhysRevD.76.083012}{Possible evidence
  for dark matter annihilations from the excess microwave emission around the
  center of the galaxy seen by the wilkinson microwave anisotropy probe}, Phys.
  Rev. D 76 (2007) 083012.
\newblock \href {http://dx.doi.org/10.1103/PhysRevD.76.083012}
  {\path{doi:10.1103/PhysRevD.76.083012}}.
\newline\urlprefix\url{https://link.aps.org/doi/10.1103/PhysRevD.76.083012}

\bibitem{Hooper:2007gi}
D.~Hooper, G.~Zaharijas, D.~P. Finkbeiner, G.~Dobler, {Prospects For Detecting
  Dark Matter With GLAST In Light Of The WMAP Haze}, Phys. Rev. D 77 (2008)
  043511.
\newblock \href {http://arxiv.org/abs/0709.3114} {\path{arXiv:0709.3114}},
  \href {http://dx.doi.org/10.1103/PhysRevD.77.043511}
  {\path{doi:10.1103/PhysRevD.77.043511}}.

\bibitem{Dobler:2007wv}
G.~Dobler, D.~P. Finkbeiner, {Extended Anomalous Foreground Emission in the
  WMAP 3-Year Data}, Astrophys. J. 680 (2008) 1222--1234.
\newblock \href {http://arxiv.org/abs/0712.1038} {\path{arXiv:0712.1038}},
  \href {http://dx.doi.org/10.1086/587862} {\path{doi:10.1086/587862}}.

\bibitem{Henry:2018yar}
R.~C. Henry, J.~Murthy, J.~Overduin, {Discovery of an Ionizing Radiation Field
  in the Universe}\href {http://arxiv.org/abs/1805.09658}
  {\path{arXiv:1805.09658}}.

\bibitem{Mattila:2019ybk}
K.~Mattila, P.~V\"ais\"anen, {Extragalactic Background Light: Inventory of
  light throughout the cosmic history}, Contemp. Phys. 60~(1) (2019) 23--44.
\newblock \href {http://arxiv.org/abs/1905.08825} {\path{arXiv:1905.08825}},
  \href {http://dx.doi.org/10.1080/00107514.2019.1586130}
  {\path{doi:10.1080/00107514.2019.1586130}}.

\end{thebibliography}

\end{document}